\documentstyle[manuscript,aps]{revtex}

\begin{document}
\draft

\title{Effect of colloidal charge discretization in the primitive model}

\author{Ren\'{e} Messina\thanks{
email: messina@mpip-mainz.mpg.de
}, Christian Holm, and Kurt Kremer}

\address{Max-Planck-Institut f\"{u}r Polymerforschung, Ackermannweg 10, 55128, Mainz,
Germany}

\date{\today{}}

\maketitle
\begin{abstract}
The effect of fixed discrete colloidal charges in the primitive model is investigated
for spherical macroions. Instead of considering a central bare charge, as it
is traditionally done, we distribute \textit{discrete} charges randomly on the
sphere. We use molecular dynamics simulations to study this effect on various
properties such as overcharging, counterion distribution and diffusion. In the
vicinity of the colloid surface the electrostatic potential may considerably
differ from the one obtained with a central charge. In the strong Coulomb coupling,
we showed that the colloidal charge discretization qualitatively influences
the counterion distribution and leads to a strong colloidal charge-counterion
pair association. However, we found that \textit{charge inversion} still persists
even if strong pair association is observed.\\

\end{abstract}

\narrowtext

\section{Introduction}

The electrostatic interactions in charged colloidal systems play a crucial role
in determining the physical properties of such materials \cite{Isralachvili_1992,Evans_book_1999}.
The behavior of these systems is extremely complex due to the \textit{long range}
Coulomb interactions. A first simplifying assumption is to treat the solvent
as a dielectric medium solely characterized by its relative permittivity \( \epsilon _{r} \).
A second widely used approximation consists in modeling the \textit{short range}
ion-ion excluded volume interaction by hard spheres. These two approximations
are the basis of the so-called primitive model of electrolyte solutions. The
system under consideration is an asymmetrical polyelectrolyte made up of highly
charged macroions and small counterions in solution. A further simplification
can be achieved by partitioning the system into subvolumes (cells), each containing
one macroion together with its neutralizing counterions plus, if present, additional
salt. This approximation has been called accordingly the cell model \cite{Hill_book_1960,Wennerstroem_JCP_1982}.
The cells assume the symmetry of the macroion, here spherical, and are electrostatically
decoupled. In this way one has reduced a complicated many-body problem to an
effective one colloid problem. For spherical macroions the structural charge
is normally modeled by a \textit{central} charge, which, by Gauss theorem, is
equivalent of considering a \textit{uniform} surface charge density as far as
the electric field \textit{outside} the sphere is concerned.

Most analytical work as well as simulation approaches rely on the above assumptions.
It is well known that in the strong Coulomb coupling regime ion-ion correlations
become very important, and significant deviations from mean-field approaches
are expected. One of the effects which the mean-field theory like Poisson-Boltzmann
can not explain is the phenomenon of overcharge, also called charge inversion.
It consists of binding excess counterions to a charged particle (macroion) so
that its net charge changes sign. This has recently attracted significant attention
\cite{Perel_Physica_1999,Shklowskii_PRE_1999b,Mateescu_EPL_1999,Joanny_EPJB_1999,Sens_PRL_1999,Marcelo_PRE_RapCom1999,Deserno_Macromol_2000,Messina_PRL_2000,Messina_EPL_2000,Nguyen_JCP_2000}.
It may give rise to a possible mechanism for strong long range attraction between
like-sign charged colloids \cite{Messina_PRL_2000,Messina_EPL_2000}.

The purpose of this paper is to investigate if such a phenomenon (overcharge)
depends on the way the structural charge is represented. The macroion is taken
to be perfectly spherical, i. e. we neglect any surface roughness \cite{Bhattacharjee_Lang_1998}.
We introduce discrete charges on the macroion sphere instead of a central charge,
and compare the results to those obtained with a central charge. We concentrate
of the following properties in the strong Coulomb coupling: overcharging, counterion
distribution and surface diffusion.

\section{Simulation model}

\subsection{Macroion charge discretization}

The macroion charge discretization is produced by using \( N_{m} \) identical
microions of diameter \( \sigma  \), all identical to the counterions, distributed
\textit{randomly} on the surface of the macroion. Then the structural charge
is \( Q=-Z_{m}e=-Z_{c}N_{m}e \), where \( Z_{m}>0 \), \( Z_{c} \) is the
counterion valency and \textit{e} is the positive elementary charge. The discrete
colloidal charges (DCC) are \textit{fixed} on the surface of the spherical macroion.
In spherical coordinates the elementary surface is given by: \begin{equation}
\label{Eq. dA}
dA=r_{0}^{2}sin\theta d\theta d\varphi =-r_{0}^{2}d(cos\theta )d\varphi \: ,
\end{equation}
 and to produce a random discrete charge distribution on the surface we generated
randomly the variables \( cos\theta  \) and \( \varphi  \). Only configurations
leading to an overlap of microions are rejected. Figure \ref{fig.setup} shows
a schematic view of the setup. Note that in real physical systems like sulfonated
latex spheres, no large heterogeneities are expected in the charge distribution,
provided that the colloid surface is relatively regular, therefore our choice
is justified. Nevertheless, the experimental situation is more complicated since
other phenomena such as surface chemical reactions \cite{Spalla_JCP_1991},
hydration, roughness \cite{Bhattacharjee_Lang_1998} and many more may be present.
Here, we restrict ourselves to a simple model in order to understand the effect
of macroion charge discretization, and leave the other questions for future
investigations.

\subsection{Molecular dynamics procedure}

We use molecular dynamics (MD) simulations to compute the motion of the counterions
coupled to a heat bath acting through a weak stochastic force \textbf{W}(t).
The equation of motion of counterion \textit{i} reads \begin{equation}
\label{eq. Langevin}
m\frac{d^{2}{\mathbf{r}}_{i}}{dt^{2}}=-\nabla _{i}U-m\Gamma \frac{d{\mathbf{r}}_{i}}{dt}+{\mathbf{W}}_{i}(t)\: ,
\end{equation}
 where \textit{m} is the counterion mass, \textit{U} is the potential force
having two contributions: the Coulomb interaction and the excluded volume interaction
and \( \Gamma  \) is the friction coefficient. Friction and stochastic force
are linked by the dissipation-fluctuation theorem \( <{{\mathbf{W}}_{i}}(t)\cdot {{\mathbf{W}}_{j}}(t')>=6m\Gamma k_{B}T\delta _{ij}\delta (t-t^{'}) \).
For the ground state simulations the fluctuation force is set to zero.

Excluded volume interactions are taken into account with a pure repulsive Lennard-Jones
potential given by \begin{equation}
\label{eq. LJ}
U_{LJ}(r)=\left\{ \begin{array}{l}
4\epsilon \left[ \left( \frac{\sigma }{r-r_{0}}\right) ^{12}-\left( \frac{\sigma }{r-r_{0}}\right) ^{6}\right] +\epsilon ,\\
0,
\end{array}\qquad \right. \begin{array}{l}
\textrm{for}\, \, r-r_{0}<r_{cut},\\
\textrm{for}\, \, r-r_{0}\geq r_{cut},
\end{array}
\end{equation}
 where \( r_{0}=0 \) for the microion-microion interaction (the microion can
be a counterion or a DCC), \( r_{0}=7\sigma  \) for the macroion-counterion
interaction, and \( r_{cut} \) (= \( 2^{1/6}\sigma  \)) is the cutoff radius.
This leads to a macroion-counterion distance of closest approach \( a=8\sigma  \).
Energy and length units in our simulations are defined as \( \epsilon = \)\( k_{B}T_{0} \)
(with \( T_{0}=298 \) K), and \( \sigma =3.57 \) \AA\ respectively.

The pair electrostatic interaction between any pair \textit{ij}, where \textit{i}
and \textit{j} denote either a DCC or a counterion, reads \begin{equation}
\label{eq. coulomb}
U_{coul}(r)=k_{B}T_{0}l_{B}\frac{Z_{i}Z_{j}}{r}\: ,
\end{equation}
 where \( l_{B}=e^{2}/4\pi \epsilon _{0}\epsilon _{r}k_{B}T_{0} \) is the Bjerrum
length describing the electrostatic strength. Being essentially interested in
the strong Coulomb coupling regime we choose the relative permittivity \( \epsilon _{r}=16 \)
(\( l_{B}=10\sigma  \)), \textit{divalent} counterions \( (Z_{c}=2) \) and
\textit{divalent} DCC for the remaining of this paper.

The macroion and the counterions are confined in a spherical impenetrable cell
of radius \textit{\( R \).} The macroion is held fixed and is located at the
center of the cell. The colloid volume fraction \textit{\( f_{m} \)} is defined
as \( r_{m}^{3}/R^{3} \), where \( r_{m}=a-\sigma /2 \) is the colloid radius.
We have fixed \( R=40\sigma  \) so that \( f_{m}=6.6\times 10^{-3} \). To
avoid image charge complications, the permittivity \( \epsilon _{r} \) is supposed
to be identical within the whole cell (including the macroion) as well as outside
the cell.

\section{Macroion electric field}

The first step to understand the effect of colloidal charge discretization consists
of estimating the electric field, or equivalently, the electrostatic \textit{potential}
generated by such a sphere in the \textit{absence} of counterions. A simple
graphical visualization of the field lines is here not possible, since there
is no perfect symmetry. Indeed, in the present case the electric field becomes
very anisotropic and irregular close to the sphere, which is the most interesting
region where correlations are expected to be large. To describe qualitatively
the effect of charge discretization on the electrostatic potential, we compute
for three perpendicular directions \textit{x}, \textit{y}, \textit{z} the resulting
\textit{radial} potentials \( V_{x}(r \)), \( V_{y}(r) \), \( V_{z}(r) \)
(see Fig. \ref{fig.setup}) for one given DCC random distribution as a function
of the distance \( r\geq a \) from the macroion center. The radial component
of the electric field \( E_{i}(r)=-\frac{\partial }{\partial r}V_{i}(r) \)
has the important feature of representing the \textit{attractive} component
towards the sphere. The normalized radial potential \( V_{i} \) in the \textit{i\( ^{th} \)}
direction at a distance \textit{\( r \)} from the colloid center is given by:

\begin{equation}
\label{eq. X-field}
V_{i}(r)=-k_{B}T_{0}l_{B}Z^{2}_{c}\sum ^{N_{m}}_{j=1}\frac{1}{|{\mathbf{r}}_{j}(r)|}\: ,
\end{equation}
 where \( {\mathbf{r}}_{j}(r) \) is the vector pointing from the microion \textit{j}
to the point where the electric potential is computed (see Fig. \ref{fig.setup}).
Physically, \( V(r) \) is the electrostatic potential interaction between a
counterion and all the surface microscopic colloid charges. The monopole contribution
is merely given by \( V_{mono}(r)=-k_{B}T_{0}l_{B}\frac{Z_{m}Z_{c}}{r} \).
In Fig. \ref{fig.colloid-surface-potential} we show the electric potential
for three typical bare charges, each corresponding to \textit{one} given random
macroion charge distribution. For all cases, one notes that at the vicinity
of the surface the potential becomes very different from the one computed with
a central charge. We carefully checked that similar results were obtained for
other choices of \textit{x}, \textit{y}, \textit{z} directions (by rotating
the trihedron (\( \mathbf{e}_{x} \), \( \mathbf{e}_{y} \), \( \mathbf{e}_{z} \))).
However if we observe the electric field sufficiently far away from the colloidal
surface (about one macroion diameter) the field is almost exactly the same as
the one produced by a central charge, which we term \textit{isotropic} for the
rest of this paper. A closer look on Fig. \ref{fig.colloid-surface-potential}
reveals that by increasing the bare charge \( Z_{m} \) the electric field starts
to become isotropic at smaller distances from the sphere's surface. This last
feature can be physically easily interpreted. In fact when one increases the
bare charge, one also increases the absolute number of discrete charges which
has the effect of approaching the uniform continuous charge density limit (corresponding
to the isotropic case).

To capture the discretization effect on the \textit{surface} electrostatic potential,
we have measured the electrostatic potential along a circle of radius \( a \)
concentric to the spherical macroion (see Fig. \ref{fig.setup}). We start from
the top of a given DCC microion and move along a circle in a random direction
and measure the electrostatic potential \( V(s) \) as a function of the arc
length separation \( s \) from the starting point. The same formula as Eq.
(\ref{eq. X-field}) has been used here. The \textit{constant} monopole contribution
is merely given by \( V_{mono}=-k_{B}T_{0}l_{B}\frac{Z_{m}Z_{c}}{a} \). Results
are reported in Fig. \ref{fig.colloid-surface-potential} for the same configurations
as before. It clearly shows that the electrostatic potential is strongly fluctuating.
More specifically, the higher the structural charge \( Z_{m} \), the larger
the {}``oscillation frequency{}'' of the potential fluctuations over the surface.
This feature can be explained in terms of {}``\textit{holes}{}''. In the very
vicinity of a given DCC the potential is increased (in absolute value) in average,
and around a given DCC there is a hole (depletion of charges) which tends to
decrease the potential (in absolute value). The average surface of this hole
is increasing with decreasing bare charge \( Z_{m} \) (i. e. decreasing density
of charged sites).

In the following sections we are going to study the effect of charge discretization
on the counterions distribution in the strong Coulomb coupling. For all following
results we used the same random charge distributions which gave the results
of Figs. \ref{fig.colloid field} and \ref{fig.colloid-surface-potential}.

\section{Ground state analysis}

In this section, we focus on counterion distribution exclusively governed by
\textit{energy minimization}, i. e. \textit{T} = 0K. In such a case correlations
are maximal, and all the counterions lie on the surface of the spherical macroion.
To avoid the trapping in metastable states, we systematically heat and cool
(10 cycles) the system and retain the lowest energy state obtained in this way.
Furthermore we choose as the starting configuration one where each DCC is exactly
associated with one counterion, and each of these \textit{dipoles} are radially
oriented (each dipole vector and the macroion center lie on the same line).
Preliminary, we checked that this method reproduces well the ground state energy
and structure in simple situations where a central charge with two, three, four
or five counterions is present. The structure of these systems is well known
by the Gillespie rules \cite{NOTE_VESPR_theory}. It turns out that in these
situations no rough energy landscape (even for \( Z_{m}=180 \) and 90 counterions)
appears and therefore the MD simulation easily finds the global minimum. It
is only in the case of DCC that several energy minima are observed.

\subsection{Neutral case}

First we consider the simple salt free case where the system {[}macroion + counterions{]}
is neutral. In order to characterize the counterion layer, we compute the counterion
correlation function (denoted by CCF) \( g(r) \) on the surface of the sphere,
defined as: \begin{equation}
\label{Eq. CCF-g(r)}
\rho _{s}^{2}g(r)=\sum _{i\neq j}\delta (r-r_{i})\delta (r-r_{j}),
\end{equation}
 where \( \rho _{s}=N_{c}/4\pi a^{2} \) is the surface counterion concentration
(\( N_{c}=Z_{m}/Z_{c} \) being the number of counterions), \textit{\( r \)}
corresponds to the arc length on the sphere. Note that at zero temperature all
equilibrium configurations are identical, thus only one is required to obtain
the CCF. Similarly, one can also define a surface macroion correlation function
(MCF) for the microions on the surface of the macroion. The CCF is normalized
as follows \begin{equation}
\label{eq. correlation-normalization}
\rho _{s}\int _{0}^{\pi a}2\pi rg(r)dr=(N_{c}-1)\: .
\end{equation}
 Because of the \textit{finite} size and the topology of the sphere, \( g(r) \)
has a cut-off at \( \pi a \) (=25.1 \( \sigma  \)). Therefore at {}``large{}''
distance the correlation function differs from the one obtained with an infinite
planar object.

The CCF and MCF for two different structural charges \( Z_{m} \) (50 and 180)
can be inspected in Fig. \ref{fig.ground-state CF}. The CCF is computed for
a system with a central charge (CC) and for the discrete colloid charges (DCC)
case. One remarks that both CCF differ considerably following the nature of
the colloidal charge, i. e., discrete or central (see Fig. \ref{fig.ground-state CF}).
For the isotropic case (central charge) a Wigner Crystal structure is observed
as was already found in Refs. \cite{Messina_PRL_2000,Messina_EPL_2000,Messina_PRE_2000}.
It turns out that when we have to deal with DCC the counterion distribution
is strongly dictated by colloidal charge distribution (see Fig. \ref{fig.ground-state CF}).
Ground state structures are depicted in Fig. \ref{fig.gs-snapshot}. It clearly
shows the \textit{ionic pairing}, between DCC and counterions. Also, it appears
natural to call such a structure a \textit{pinned} configuration. However one
can expect that the structure might become less pinned if the typical intra-dipole
distance (here \( \sigma  \)) and the typical mean inter-dipole distance become
of the same order. This is a case which is not discussed in the present paper.
It would correspond to extremely highly charged colloids that are rarely encountered
in nature. Nevertheless, we checked that even for \( Z_{m}=360 \) the structure
is still pinned, where the average inter-dipole distance is about \( 2\sigma  \).

\subsection{Overcharge}

We now investigate the overcharge phenomenon. The starting configurations corresponds
to neutral ground states as were previously obtained. The spirit of this study
is very similar to the one undertaken in Ref. \cite{Messina_PRL_2000}. To produce
overcharge, one adds successively overcharging counterions (OC) at the vicinity
of the colloidal surface. Thus the resulting system is no longer neutral. By
using Wigner crystal theory \cite{Shklowskii_PRE_1999b,Bonsall_PRB_1977}, we
showed that the gain in electrostatic energy (compared to the neutral state)
by overcharging a single \textit{uniformly} charged colloid can be written in
the following way \cite{Messina_PRL_2000,Messina_EPL_2000,Messina_PRE_2000}:\begin{equation}
\label{Eq.oc}
\Delta E^{OC}_{n}=\Delta E^{cor}+\Delta E^{mon}=-n\gamma \sqrt{N_{c}}\left[ \frac{3}{2}+\frac{3n}{8N_{c}}\right] +(k_{B}T_{0})l_{B}Z_{c}^{2}\frac{(n-1)n}{2a}\: ,
\end{equation}
 where \( \Delta E^{cor} \), which is equal to the first term of the right
member, denotes the gain in energy due to ionic correlations for \( n \) OC.
The functional form of this term can be derived from WC theory \cite{Messina_PRL_2000,Messina_EPL_2000,Messina_PRE_2000}.
The second term on the right hand side, \( \Delta E^{mon} \), is the monopole
repulsion, which sets in when the system is overcharged (with \( n>1 \)). This
term will, for sufficient high number \textit{\( n \)} of OC, stops the process
of overcharging. As before \( N_{c}=Z_{m}/Z_{c} \) is the number of counterions
in the neutral state, and \( \gamma  \) is a constant which was determined
by using the measured value of \( \Delta E^{OC}_{1} \) of our simulations.

The total electrostatic energy of the system as a function of the number of
OC is displayed Fig. \ref{fig.gs-OC} for four bare charges \( Z_{m} \) (50,
90, 180 and 360 ). These energy curves corresponding to discrete systems were
produced by averaging over five random DCC realizations. Again, the overcharging
process is affected by the charge discretization and pinning, but it is still
energetically \textit{favorable}. The main effects of charge discretization
are: (i) the reduction of gain of energy and (ii) the reduction of maximal (critical)
acceptance of OC. Both points can be explained in terms of \textit{ion-dipole}
interaction. It is exactly this \textit{attractive} ion-dipole correlations
which is responsible of charge inversion for colloidal systems with discrete
charges. When the first OC is present, it is normally located in between the
pinning centers, and will essentially interact with its neighboring dipoles
(DCC-counterion). This interaction increases with decreasing OC-dipole separation,
i. e. increasing colloid bare charge \( Z_{m} \). This explains why the energy
gained increases with \( Z_{m} \) (see Fig. \ref{fig.gs-OC}). On the other,
the repulsion between the counterions is not fully minimized since they do not
adopt the ideal Wigner crystal structure that is obtained with a central charge
which in turn explains (i). For a higher degree of overcharge, one has to take
into account a repulsive monopolar contribution identical to \( \Delta E^{mon} \)
appearing in Eq. (\ref{Eq.oc}). Again, since for DCC structures counterions
are not perfectly ordered, the attractive correlational energy is smaller (in
absolute value) than the one obtained with a central charge, which in turn explains
(ii). Note that for very high bare charge (\( Z_{m}=360 \)) the overcharge
curve obtained with DCC approaches the one from the continuous case as expected.

Common features of overcharging between isotropic and discrete systems are briefly
given here. We note that the maximal (critical) acceptance of OC (4, 6 and 8
for a central charge and 2,4 and 6 for DCC) increases with the macroionic charge
\( Z_{m} \) (50, 90 and 180 respectively). Furthermore, for a given number
of OC, the gain in energy is always increasing with \( Z_{m} \). Also, for
a given macroionic charge, the gain in energy between two successive overcharged
states is decreasing with the number of OC. Note that at \textit{T} = 0, the
value \( \epsilon _{r} \) acts only as a prefactor. All these features are
captured by Eq. (\ref{Eq.oc}).

\section{Finite temperature}

In this part, the system is brought to room temperature \( T_{0} \). We are
interested in determining the counterions distribution as well as the counterion
motion within the counterion layer. The radius \( R \) of cell is again fixed
to \( 40\sigma  \) so that the macroion volume fraction \( f_{m} \) has the
\textit{finite} value \( 6.6\times 10^{-3} \). Under these conditions the system
is still highly energy dominated so that at equilibrium all counterions lie
on the surface of the macroion (strong condensation).

\subsection{Counterions distribution}

Like in the ground state analysis, we characterize the counterion distribution
via its surface correlation function. At non zero temperature, correlation functions
are computed by averaging \( \sum _{i\neq j}\delta (r-r_{i})\delta (r-r_{j}) \)
over 1000 independent equilibrium configurations which are statistically uncorrelated.
Results are depicted in Fig. \ref{fig.temperature-CF}. For both bare charges
\( Z_{m} \) (50 and 180) considered the counterions distribution are affected
by charge discretization. The effect of temperature is to smooth the CCF. As
expected, for the central charge case, the counterion positional order is much
weaker at room temperature than in the ground state case.

\subsection{Surface diffusion}

The aim of this section is to answer the following question: do the counterions
only oscillate around their equilibrium (ground state) position or do they have
also a global translational motion over the sphere?

To answer to this question one introduces the following quantity: \begin{equation}
\label{eq. MSD}
\Delta x^{2}(t,t_{0})=\frac{1}{t-t_{0}}\int ^{t}_{t_{0}}dt^{'}[x(t^{'})-x(t_{0})]^{2}\: ,
\end{equation}
 which is referred as the \textit{mean square displacement} (MSD), where \( x(t_{0}) \)
represents the position of a given counterion at time \( t=t_{0} \) and \( x(t,t_{0}) \)
is its position at later time \textit{t}. The root mean square displacement
(RMSD) is defined as \( \Delta x(t,t_{0})=\sqrt{\Delta x^{2}(t,t_{0})} \).
Like for the surface correlation function, the MSD is measured on the spherical
surface (arc length) and it has a natural upper limit \( (\pi a)^{2} \). Results
for the discrete case are depicted in Fig. \ref{fig.surface diffusion} for
two macroion bare charges \( Z_{m} \) (50 and 180), where each counterion'
RMSD is given. In both cases, the RMSD is smaller than the typical mean inter-dipole
separation, which is approximatively \( (\rho _{s})^{-1/2}\approx 6\sigma  \)
for \( Z_{m}=50 \) and \( (\rho _{s})^{-1/2}\approx 3\sigma  \) for \( Z_{m}=180 \).
This suggests that the motion of the counterion is \textit{purely} \textit{oscillatory}
around its DCC pinning center. Fig. \ref{fig.surface diffusion} also shows
that pinning is stronger for \( Z_{m}=50 \) than for \( Z_{m}=180 \). This
is in agreement with our previous statement, where we point out that the inter-dipole
distance has to be comparable (or smaller) to (than) the intra-dipole distance
in order to reduce pinning effect. Thus the DCC sites do produce a noticeable
energy well. One can get convinced on this point, by evaluating the electrostatic
binding energy of an ionic pair \( E_{pin} \) which is \begin{equation}
\label{eq.pinning-energy}
E_{pin}=-k_{B}T_{0}l_{B}Z^{2}_{c}/\sigma =-40k_{B}T_{0}.
\end{equation}
 However, for much higher temperatures a liquid like behavior is to be expected.
Also, of course, the strength of the pinning can be lowered by different parameters:
larger ions, higher dielectric constant \( \epsilon _{r} \), monovalent ions
as it is captured by Eq. (\ref{eq.pinning-energy}). For the isotropic case,
we have checked that counterions have a large lateral motion and can move all
over the sphere. This is obvious since in this situation there are no pinning
centers.

\section{Concluding remarks}

We have carried out MD simulations within the framework of the primitive model
to elucidate the effect of colloidal charge discretization . The main result
of our study is that, in the strong Coulomb coupling, the charge inversion is
still effective when the macroion structural charge is carried by discrete charges
distributed randomly over its surface area. We have shown that the intrinsic
electrostatic potential solely due to the surface colloidal microions strongly
vary from point to point on the macroion sphere. When counterions are present,
it leads to a pinned structure where every counterion is associated with one
colloidal charge site. Furthermore we observed a pure phonon-like behavior (counterion
vibration with small lateral motion) is found at room temperature.

Future work will address the problem of valency asymmetry, that is when the
colloidal charges are represented by monovalent counterions and the counterions
are divalent. This is a non trivial situation since ionic pairing may be frustrated.
Also, the case of the low Coulomb coupling regime should be addressed.

\acknowledgments
We thank B. Shklovskii for helpful and constructive remarks. This work is supported
by \textit{Laboratoires Europ\'{e}ens Associ\'{e}s} (LEA).

\begin{figure}

\caption{Schematic view of the setup: the discrete colloidal charges (DCC) of diameter
\protect\( \sigma \protect \) are in dark grey. The radial electrostatic field
components \protect\( E_{x}\protect \) and \protect\( E_{y}\protect \) are
represented. For a detailed meaning of the other symbols see text. Note that
this a a two-dimensional representation of the three-dimensional system. }

\label{fig.setup}
\end{figure}

\begin{figure}

\caption{Radial electrostatic potential as a function of macroion center distance \protect\( r\protect \)
produced by the fixed microscopic colloidal charges disposed on the sphere.
These potential have been measured in three perpendicular directions (x, y,
z) directions (see Fig. \ref{fig.setup}). The isotropic case corresponds to
the field obtained with a central charge (monopole). Three structural charges
are considered: (a) \textit{\protect\( Z_{m}=50\protect \)} (b) \textit{\protect\( Z_{m}=90\protect \)}
and (c) \textit{\protect\( Z_{m}=180\protect \)}. }

\label{fig.colloid field}
\end{figure}

\begin{figure}

\caption{Surface electrostatic potential as a function of the arc length \protect\( s\protect \)
along a circle of radius \protect\( a\protect \) concentric to the the macroion
for three different trajectories. The monopole contribution is represented by
the dashed line. The same configurations as those of Fig. \ref{fig.colloid field}(a-c)
have been used. }

\label{fig.colloid-surface-potential}
\end{figure}

\begin{figure}

\caption{Ground state surface correlation functions for two macroion bare charges (a)
\textit{\protect\( Z_{m}=50\protect \)} and (b) \textit{\protect\( Z_{m}=180\protect \)}.
The two counterion correlation functions (CCF) are obtained for discrete colloidal
charges (DCC) and for the central charge (CC). To get the same distance range
for CCF and the colloidal surface discrete microions correlation function (MCF),
the MCF curve x-axis (\protect\( r/\sigma \protect \)) was rescaled by a factor
\protect\( a/r_{0}\protect \) (compare setup Fig. \ref{fig.setup}). }

\label{fig.ground-state CF}
\end{figure}

\begin{figure}

\caption{Ground state structures for two values (a)\textit{ \protect\( Z_{m}=50\protect \)}
and (b) \protect\( Z_{m}=180\protect \) corresponding to the two cases of Fig.
\ref{fig.ground-state CF}. The colloidal surface microions are in white, and
the counterions in blue. Full ionic pairing association occurs.}

\label{fig.gs-snapshot}
\end{figure}

\begin{figure}

\caption{Electrostatic energy (in units of \protect\( k_{B}T_{0}\protect \)) for ground
state configurations of a single charged macroion as a function of the number
of \textit{overcharging} counterions for three different bare charges \textit{\protect\( Z_{m}\protect \)}.
CC stands for the central charge case. The neutral case was chosen as the potential
energy origin. Dashed lines are produced by using equation (\ref{Eq.oc}). }

\label{fig.gs-OC}
\end{figure}

\begin{figure}

\caption{Surface correlation functions at \textit{room temperature}. The two CCF are
obtained for discrete colloidal charges (DCC) and for the central charge (CC).
(a) \textit{\protect\( Z_{m}=50\protect \)} (b) \textit{\protect\( Z_{m}=180\protect \)}. }

\label{fig.temperature-CF}
\end{figure}

\begin{figure}

\caption{Root mean square displacement (rmsd) for each counterion. (a) \textit{\protect\( Z_{m}=50\protect \)}
(b) \textit{\protect\( Z_{m}=180\protect \)}. }

\label{fig.surface diffusion}
\end{figure}


\begin{thebibliography}{10}
\bibitem{Isralachvili_1992} J. Israelachvili, \emph{Intermolecular and Surface Forces} (Academic, London,
1992).
\bibitem{Evans_book_1999}D.~F. Evans and H. Wennerstr\"{o}m, \emph{The Colloidal Domain} (Wiley-VCH,
New York, 1999).
\bibitem{Hill_book_1960} T.~L. Hill, \emph{Statistical mechanics} (Addison-Wesley, Reading, Mass.,
1960).
\bibitem{Wennerstroem_JCP_1982}H. Wennerstr\"{o}m, B. J\"{o}nsson, and P. Linse, J. Chem. Phys. \textbf{76},
4665 (1982).
\bibitem{Perel_Physica_1999} V. Perel and B. Shklovskii, Physica \textbf{274A}, 446 (1999).
\bibitem{Shklowskii_PRE_1999b} B. Shklovskii, Phys. Rev. E \textbf{60}, 5802 (1999).
\bibitem{Mateescu_EPL_1999} E.~M. Mateescu, C. Jeppesen, and P. Pincus, Europhys. Lett. \textbf{46}, 493
(1999).
\bibitem{Joanny_EPJB_1999} J.~F. Joanny, Europ. J. Phys. B \textbf{9}, 117 (1999).
\bibitem{Sens_PRL_1999} E. Gurovitch and P. Sens, Phys. Rev. Lett. \textbf{82}, 339 (1999).
\bibitem{Marcelo_PRE_RapCom1999} M. Lozada-Cassou, E. Gonz\'{a}lez-Tovar, and W. Olivares, Phys. Rev. E \textbf{60},
R17 (1999).
\bibitem{Deserno_Macromol_2000} M. Deserno, C. Holm, and S. May, Macromolecules \textbf{33}, 199 (2000).
\bibitem{Messina_PRL_2000} R. Messina, C. Holm, and K. Kremer, Phys. Rev. Lett. \textbf{85}, 872 (2000).
\bibitem{Messina_EPL_2000}R. Messina, C. Holm, and K. Kremer, Europhys. Lett. \textbf{51}, 461 (2000).
\bibitem{Nguyen_JCP_2000} T.~T. Nguyen, A.~Y. Grosberg, and B.~I. Shklovskii, J. Chem. Phys. \textbf{113},
1110 (2000).
\bibitem{Bhattacharjee_Lang_1998} S. Bhattacharjee, C.~H. Ko, and M. Elimelech, Langmuir \textbf{14}, 3365 (1998).
\bibitem{Spalla_JCP_1991} O. Spalla and L. Belloni, J. Chem. Phys. \textbf{95}, 7689 (1991).
\bibitem{NOTE_VESPR_theory}A theory that predicts molecular geometries using the notion that valence electron
pairs occupy sites around a central atom in such a way as to minimize electron-pair
repulsion. See for example D. W. Oxtoby, H. P. Gillis and N. H. Nachtrieb, \textit{Principles
of Modern Chemistry} (Saunders College Publishing, 1999), Chap. 3, p. 80.
\bibitem{Messina_PRE_2000}R. Messina, C. Holm, and K. Kremer, submitted.
\bibitem{Bonsall_PRB_1977} L. Bonsall and A.~A. Maradudin, Phys. Rev. B \textbf{15}, 1959 (1977).
\end{thebibliography}
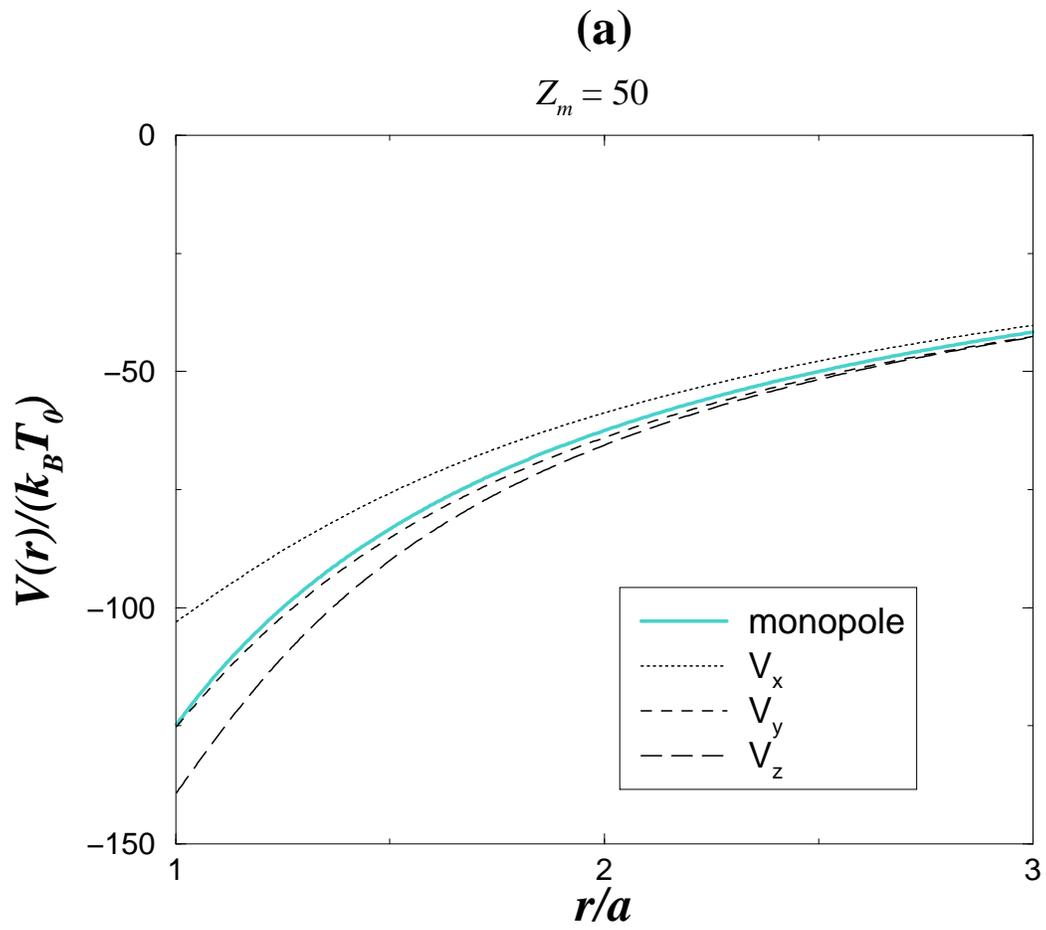
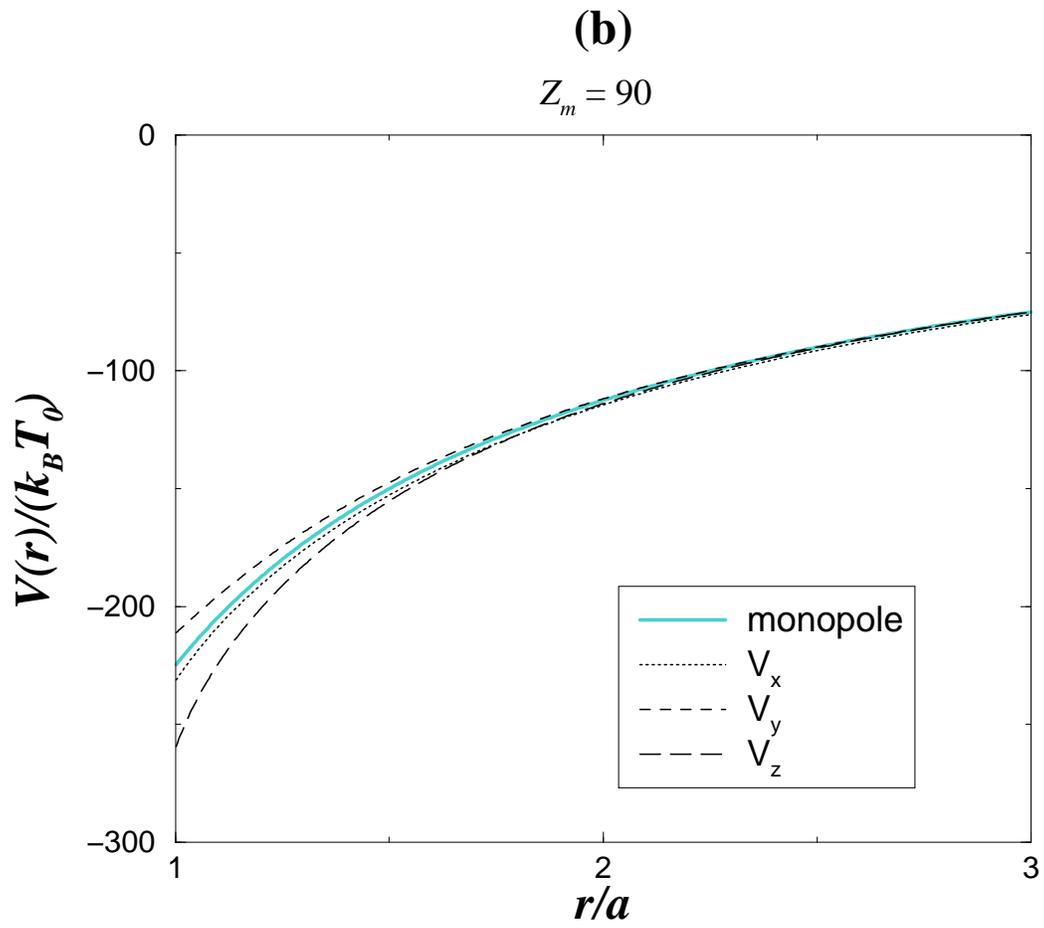
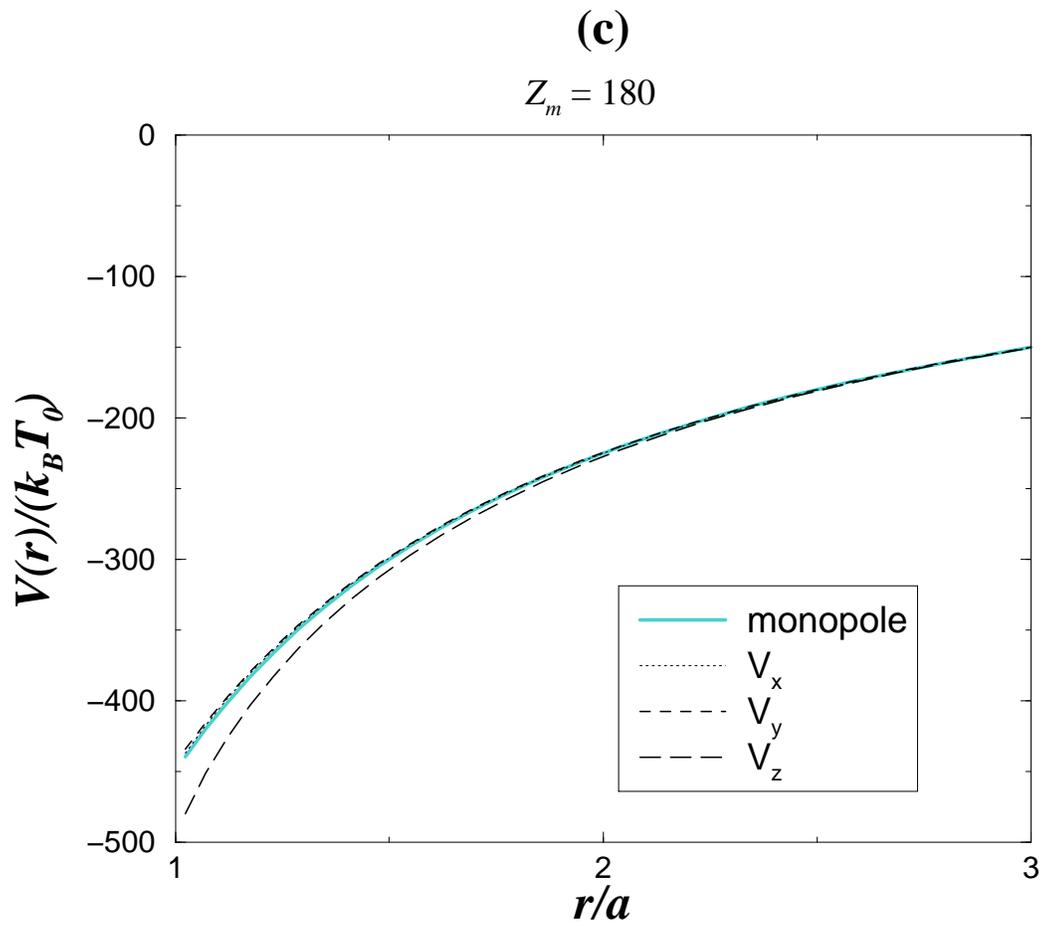
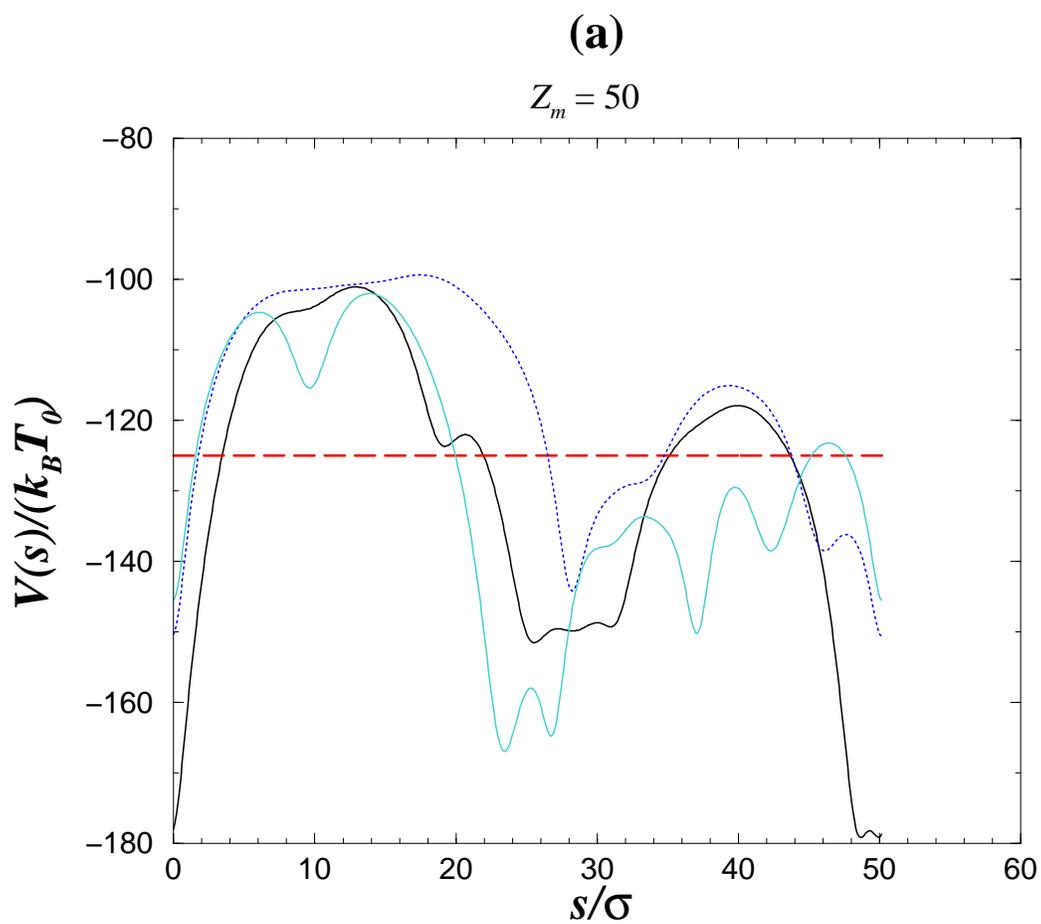
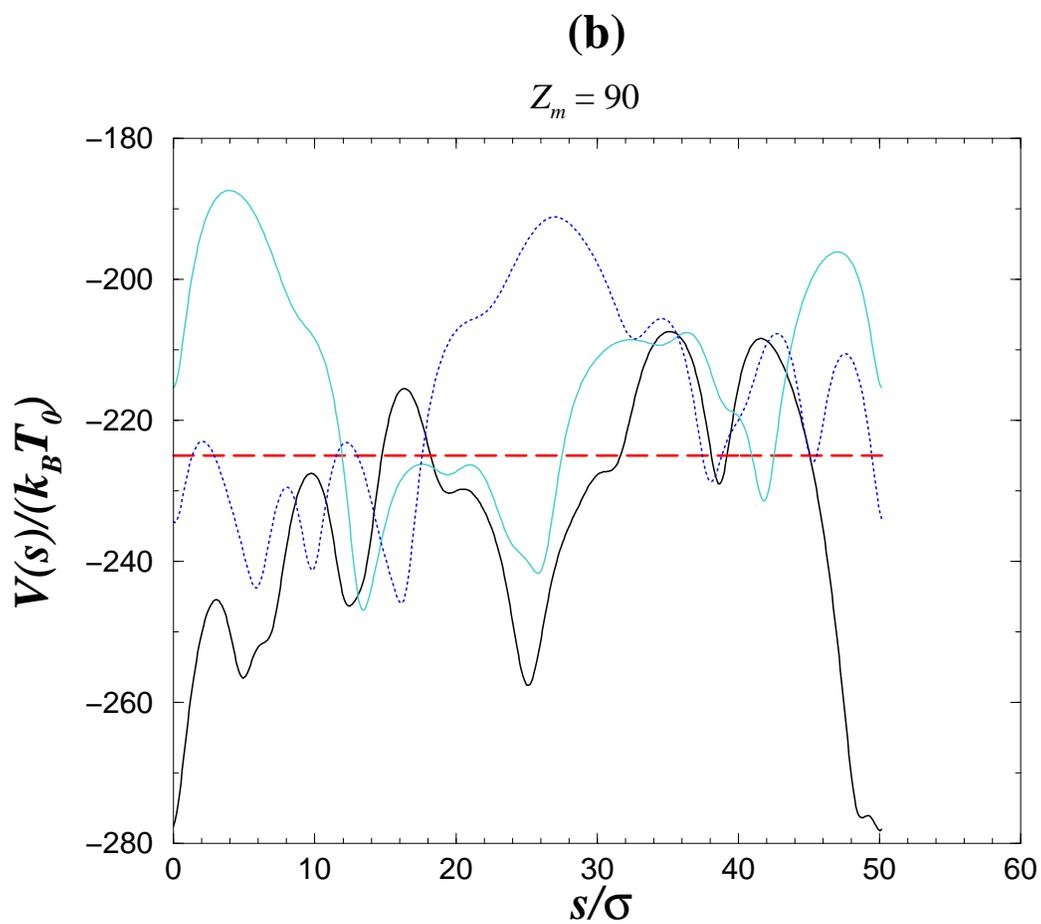
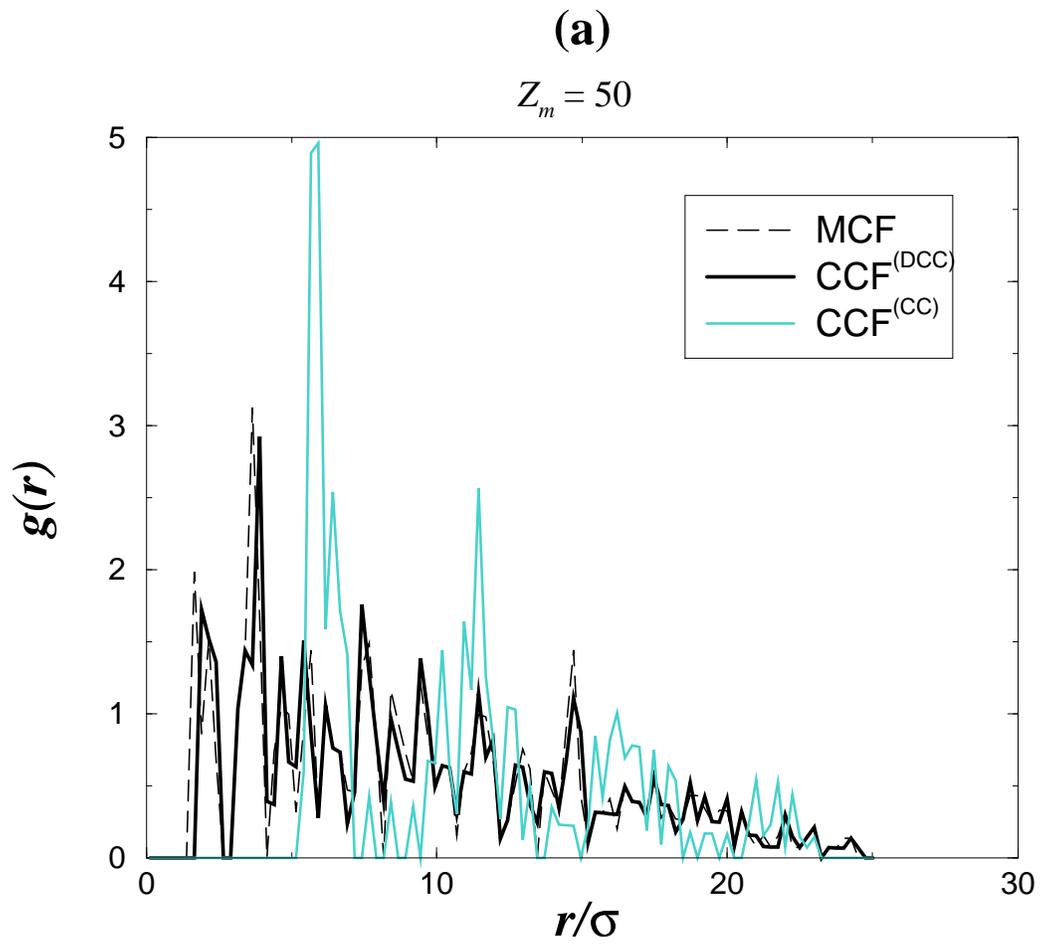
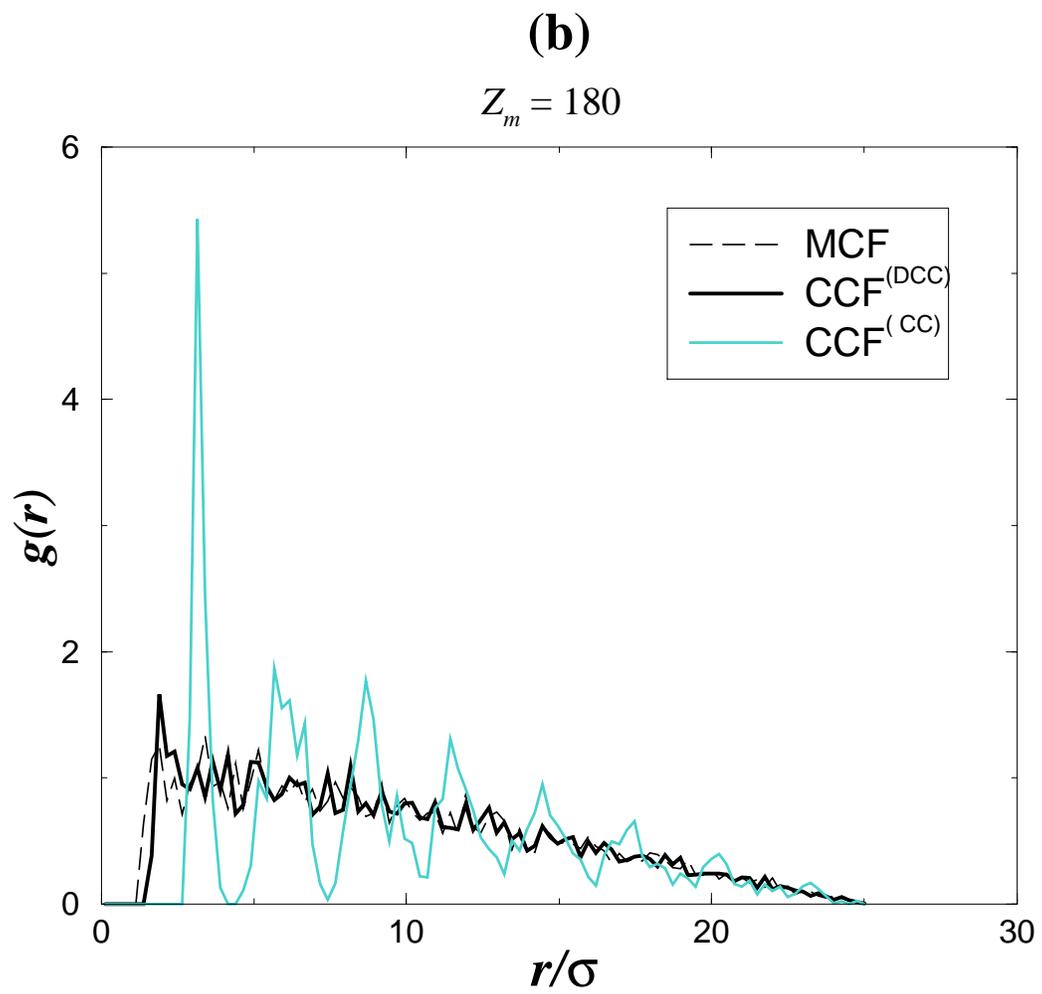
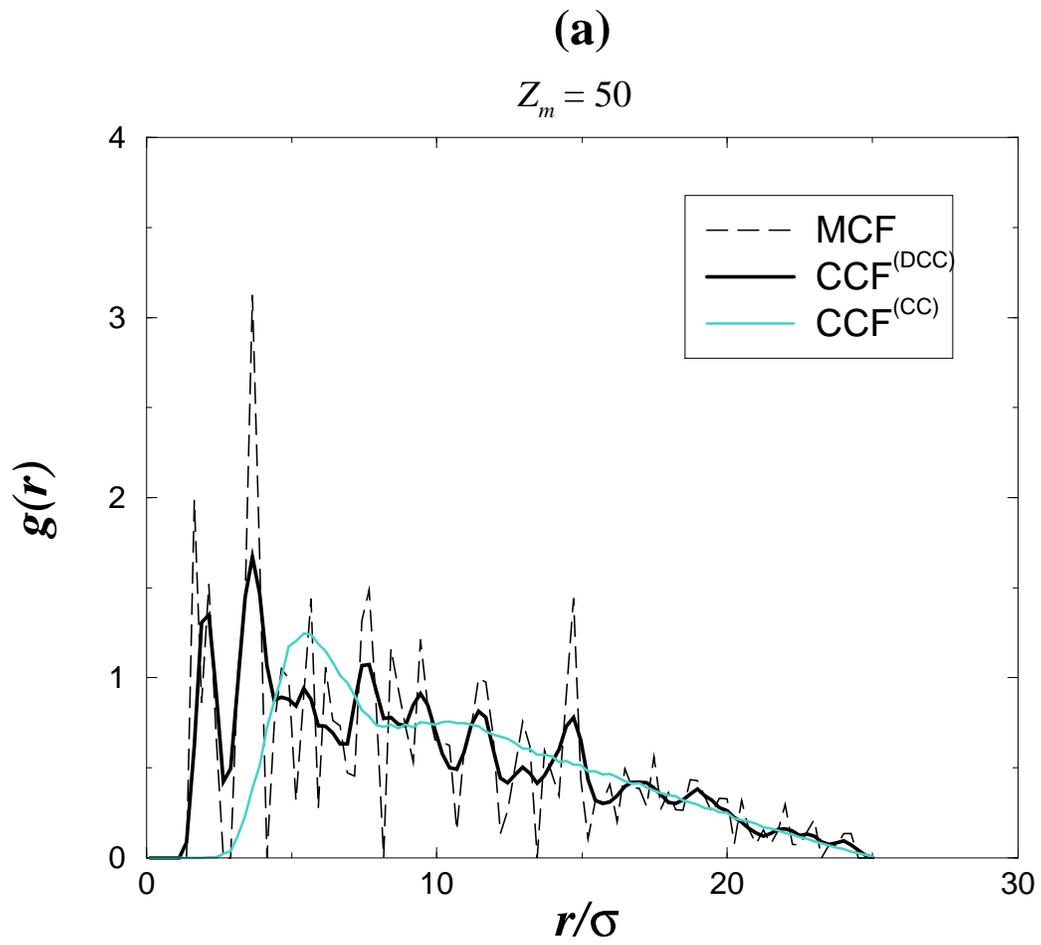
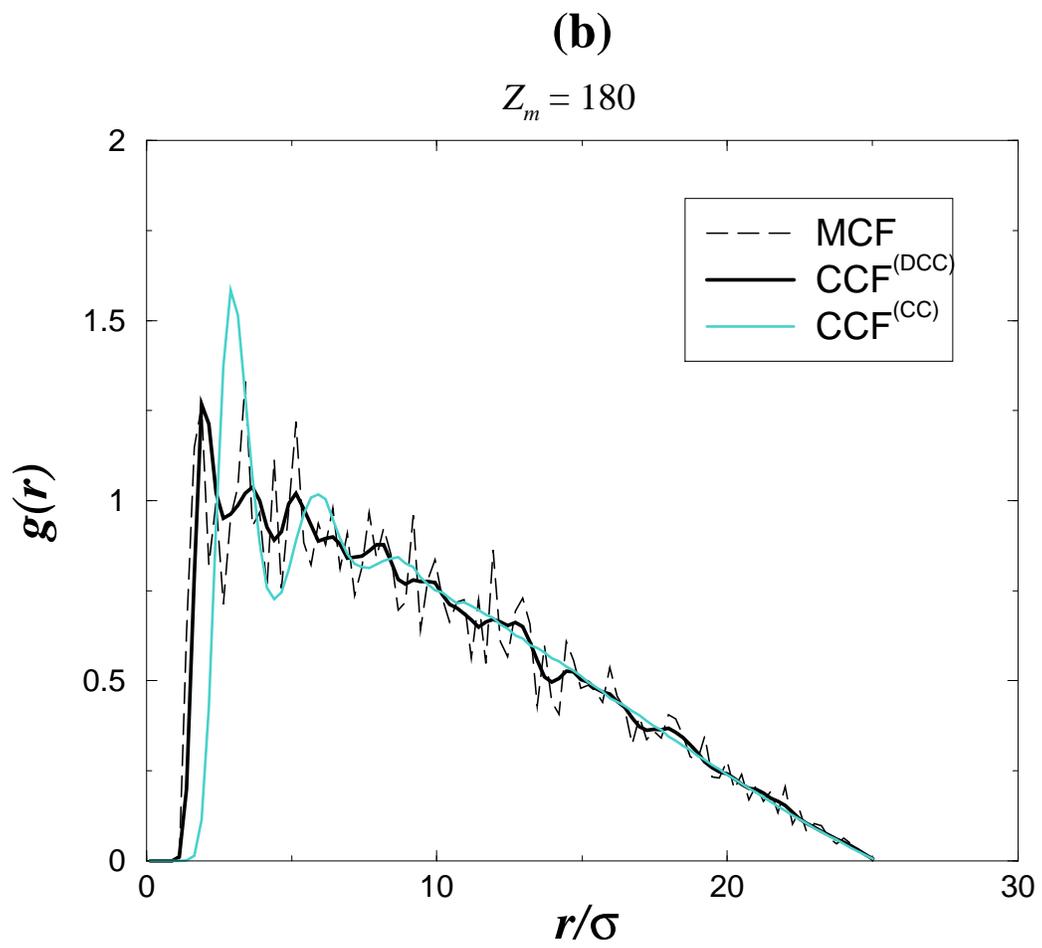
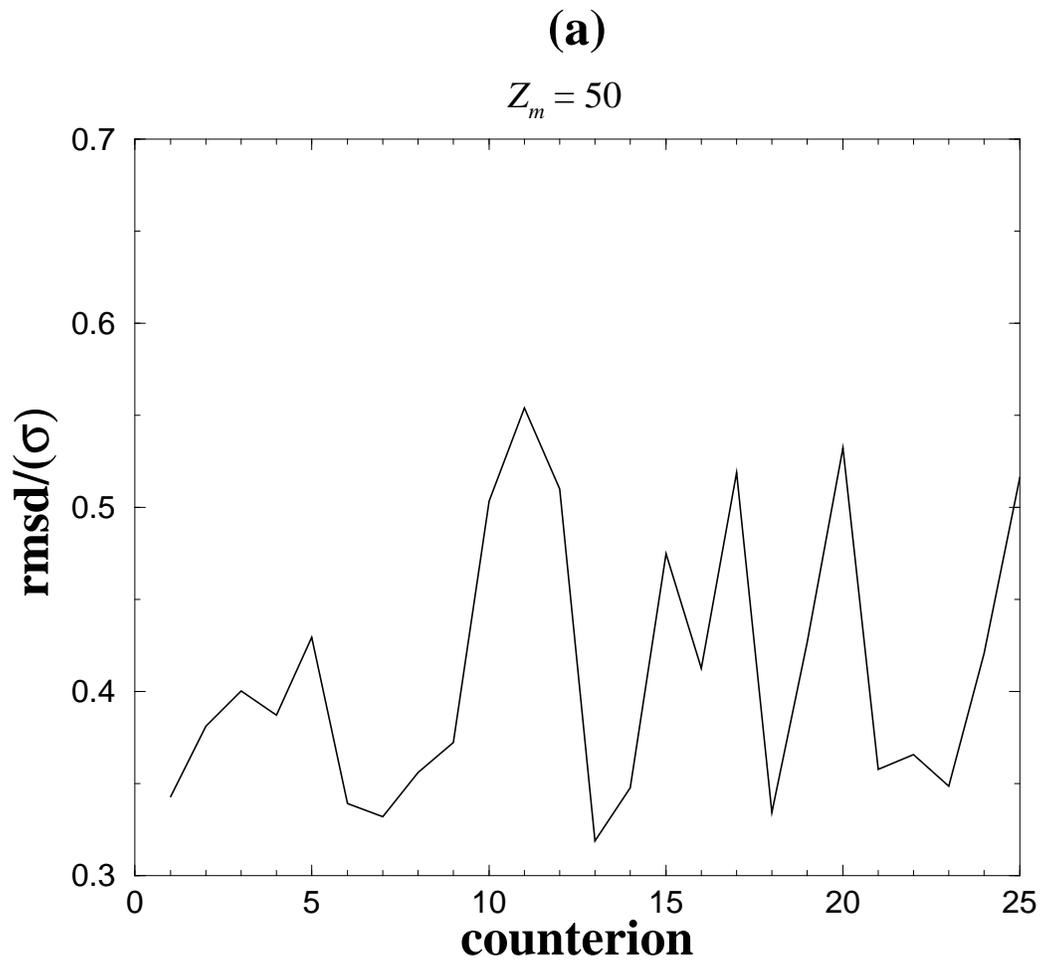
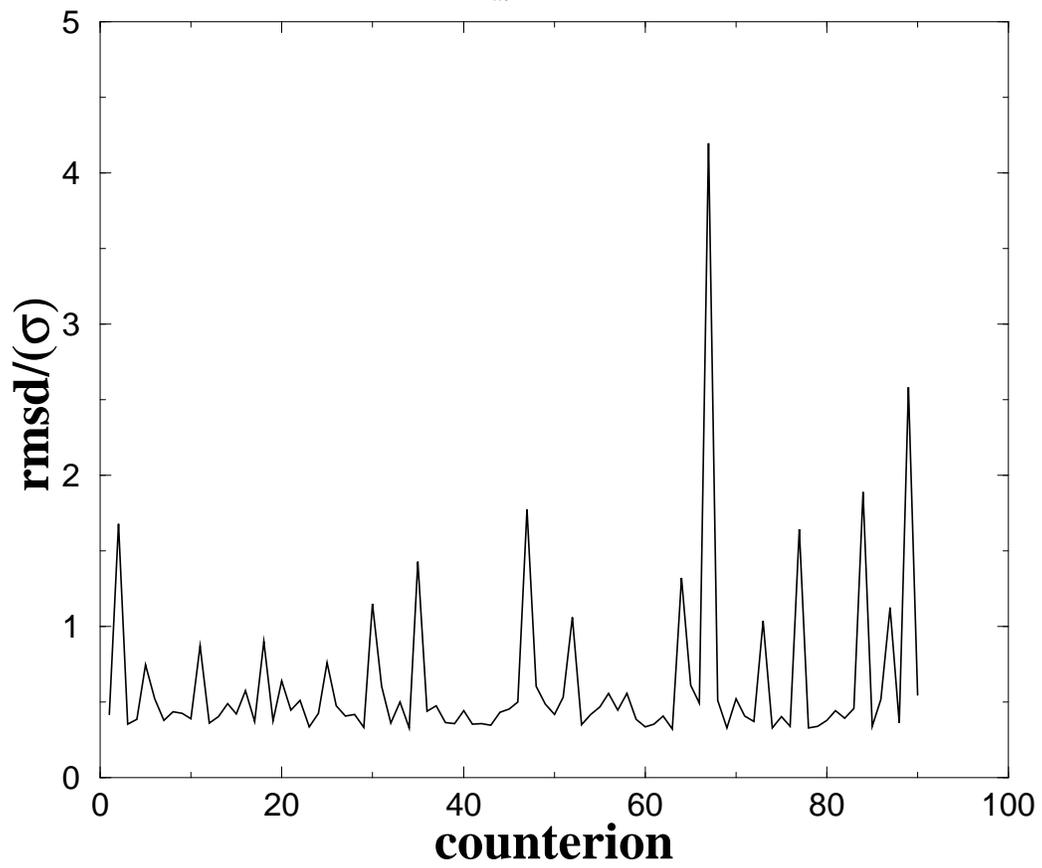
\end{document}